# Gate Stack Dielectric Degradation of Rare-Earth Oxides Grown on High Mobility Ge Substrates

*Md. Shahinur Rahman[a],[b], E. K. Evangelou[b], N. Konofaos[c], and A. Dimoulas[d]*

[a] *Detector Laboratory-GSI Helmholtzzentrum Schwierionenforschung, 64291-Darmstadt, Germany*

[b] *Lab of Electronics-Telecoms & Applications, Department of Physics, University of Ioannina, 45110-Ioannina, Greece.*

[c] *Department of Informatics, Aristotle University of Thessaloniki , Greece,*

[d] *MBE Lab. Institute of Materials Sciences, Demokritos, 15310-Athens, Greece.*

## Abstract

We report on the dielectric degradation of Rare-Earth Oxides (REOs), when used as interfacial buffer layers together with $HfO_2$ high-*k* films (REOs/$HfO_2$) on high mobility Ge substrates. Metal-Oxide-Semiconductor (MOS) devices with these stacks, show dissimilar charge trapping phenomena under varying levels of Constant-Voltage-Stress (CVS) conditions, which also influences the measured densities of the interface ($N_{it}$) and border ($N_{BT}$) traps. In the present study we also report on $C$-$V_g$ hysteresis curves related to $N_{it}$ and $N_{BT}$. We also propose a new model based on Maxwell-Wagner instabilities mechanism that explains the dielectric degradations (current decay transient behavior) of the gate stack devices grown on high mobility substrates under CVS bias from low to higher fields, and which is unlike to those used for other MOS devices. Finally, the time dependent degradation of the corresponding devices revealed an initial current decay due to relaxation, followed by charge trapping and generation of stress-induced leakage which eventually lead to hard breakdown after long CVS stressing.

---

[a] corresponding authors' email: M.S.Rahman@gsi.de

# 1. Introduction

Rate-earth Oxides (REOs) such as $CeO_2$, $Dy_2O_3$, $La_2O_3$ are used for the construction of gate stacks on Germanium substrates demonstrating excellent passivation of the surface and electrical properties[1]. However, it is important to clarify a number of reliability concerns such as, charge migration at the interface of the two dielectrics, charge trapping inside the bulk of the oxides, defects generation under bias, stress-induced leakage current (SILC), and oxides degradation issues. One of the serious drawbacks in most of the high-*k* dielectrics is the charge trapping in the bulk of the oxides. This precludes accurate extraction of mobility, flatband (threshold) voltage shift ($V_{FB/th}$) while it will also result in the degradation of the device electrical characteristics. When the MOS capacitors, (MOSCAPs) are stressed under pulses of constant gate voltage, the flatband voltage ($V_{FB}$) is extracted from Capacitance-Voltage ($C$-$V_g$) or Current-Voltage ($J_g$-$V_g$) measurements taken between the pulses. This is monitored as a function of either the stress time or the injected charge while the leakage current is recorded simultaneously. During the measurements, a good setup is maintained to avoid the influences of external charging/discharging effects to the built-in defects in the dielectrics itself. It has to be noted here that even little $C$-$V_g$ (or $J_g$-$V_g$) hysteresis can significantly affect the outcome (instability) [2].

The bulk or interfacial defects give rise to transient gate currents while it has widely been accepted that the defects existing in high-$k$ dielectrics play an important role when the devices are in operation[3]. Moreover, during a stress bias, new neutral defects/traps are created in the oxides. Depending on the stress conditions this creation of the new defects affects the resultant external leakage current. Stress-induced leakage current (SILC) is the signature of the defect generation within the gate stacks, and it is independent of the dielectrics in the stacks. The continuation of the charge trapping and defect generation leads to dielectric degradation and eventually causes hard breakdown (HBD) of the devices[3]. SILC is not only related to the generation of new defects but results from the localized, defect related weak spots near the injecting interface[4-5]. The newly generated traps are uniformly distributed in the bulk of the oxides. Moreover, the interface traps play a crucial role on the dielectric degradation and the electrical instabilities[6]. In the high-$k$ dielectrics the gate stack itself causes charge accumulation and exhibits a decay current transient behavior. This current decay behavior can be explained as Maxwell-Wagner (M-W) instability [7], named as one of the major drawbacks of gate stacks technology since it hinters the passivation quality of the high mobility substrates surface, e.g. Ge.

Another class of oxides defects, were introduced and termed by Fleetwood as "Border traps", or "Near interface oxide traps, NOIT" [8]. These border traps exchange charge with the semiconductor substrate on the time scale of the measurements being performed [9]. This charge exchange is typically slower than that for interface traps, so sometimes these defects are called "slow states". As the MOS devices continue to scale down rigorously, the influences of the interface and the border traps on device performance and reliability become more important.

When charges are accumulated at the interface of the bilayer dielectrics, this situation accelerates the relaxation polarization due to the different conductivities of the dielectric materials. Dielectric relaxation follows the direction of the applied external voltage gradient ($dV_g/dt$) when devices are under bias (CVS). The simultaneous effects of the charge migration at the interface and the relaxation polarization of the multilayer gate stacks cause the current decay transient which may termed as Maxwell-Wagner instabilities (M-W) [7]. This relaxation behaviour is observed in very low level external circuit current regime, however at medium or higher bias the situation is different and it includes charge trapping and creation of new neutral defects in the bulk of oxides. When an external field is applied across a film, it separates the bound charges, thus resulting in polarization and a compensating internal field.

In this paper we are dealing with the electrical reliability characteristics of rare-earth oxides based gate stacks (REOs-HfO$_2$) step by step for the result of long CVS dielectric degradation of Ge base MOS devices.

## 2. Experimental

Thin films of REOs/HfO$_2$ oxide stacks were prepared by molecular beam deposition (MBD) on both *p*- and *n*-type Ge (100) substrates. The REO used were Dy$_2$O$_3$, La$_2$O$_3$, and CeO$_2$. Native oxide was desorbed in situ under ultra high vacuum (UHV) conditions by heating the substrate to 225-360 °C for 15 min until a (2 × 1) reconstruction appears in the (RHEED) pattern, indicating a clean (100) surface. Subsequently, the substrate was cooled down to 225-336 $^0$C, where the oxide stacks were deposited. The surface was exposed to atomic O beams generated by a radio-frequency plasma source with simultaneous e-beam evaporation of RE/Hf at a rate of about ~0.15 Å/s. Metal–insulator–semiconductor capacitors were prepared by shadow mask and e-beam evaporation of 30-nm-thick Pt electrodes to define circular dots of 200-800 μm in diameter. The back ohmic contact was made using eutectic InGa alloy. The devices were subjected to electrical stress under CVS conditions at accumulation [1010] using a Keithley 617 source/meter, and the same instrument was used for measurements of the current for successive stress cycles versus time ($J_g-t$) and the current–voltage ($J_g-V_g$) curves. The $C-V_g$ curves at high frequency (100 kHz) were measured with an Agilent 4284A LCR meter [7, 10]. All measurements were performed in a dark box and at room temperature.

## 3. Results and discussions:

### i) $C-V_g$ characteristics of Ge-based devices: Anomalous trapping behavior

Fig. 1 shows the $C–V_g$ curves for low and medium CVS (-2V and -3V respectively) on MOS devices structured as: Pt/Dy$_2$O$_3$/$p$-Ge. The curves are recorded on fresh samples under forward and reverse bias sweeps with a gate voltage sweep rate of 50mVs$^{-1}$, and after 10 successive stresses of 500s each (i.e. $t_{stress}$=5000s), for the shake of clarity only fresh and after 5000s CVS biased $C-V_g$ curves are plotted in Fig.1.

The flatband voltage shift ($\Delta V_{FB}$), which is related to the charge trapping in the devices, show both positive and negative $\Delta V_{FB}$ shifts at the same bias polarity. In the present case, the $p$-type Ge substrates supply holes at accumulation and we expect hole trapping (see Fig.1b) in the dielectrics uniformly distributed in the bulk of the oxide or at the interface. However, Fig.1a depicts the unusual nature charge trapping characteristics on the $C$-$V_g$ curves. This could be attributed to the fact that the gate bias is always negative at accumulation hence electrons from the gate are injected into the dielectric and captured by the preexisting traps. Nevertheless the applied field ($E_{Dy2O3}$ ~2MV/cm at $t_{stress}$=0s) is not high enough to force the electrons to escape from the defects and drive them towards the substrate.

Structural measurements, e.g. TEM and XRR [11], showed that an additional ultrathin layer of $GeO_2$ was formed as an interfacial layer (*il*) with a lower-medium value of dielectric constant ($k$~5) which finally worked as a gate stack structure. The aforementioned unalike charge trapping or the change of sign of charge trapping, at accumulation, could also be happening due to the dissimilar conductivities of the bilayer insulating films, by causing a switching of the trapped charge sign by varying the gate voltage [12], or could be relaxation behavior in the gate stacks [15]. Recently published work also suggested that this nature of charge trapping is, due to relaxation and Maxwell-Wagner instabilities (M-W) [7, 13].

The typical trapping phenomenon has also been observed for other REOs used in gate stack MOS devices [7, 10], and at the current moment it is not well-understood. We also emphasize here that a large hysteresis of the $C$-$V_g$ curves (see Fig. 1) is observed for Ge based MOS devices [11, 14-15], due to either the intermixing of the high-$k$ and the interfacial layer, or to an excess amount of positive charges in the bulk of the dielectric of the gate stacks.

**ii) Border traps characteristics in REOs and its gate stacks**

Our previous results, e.g. TEM and XPS, [16] showed that when $CeO_2$ was directly deposited on high mobility Ge substrates, it interacted strongly with the substrate and spontaneously formed a 1~2nm thick interfacial layer thus leading to a gate stack structure. Fig. 2 shows the $C$-$V_g$ curve (circled symbol-line) and its hysteresis characteristics (solid line) of a $CeO_2$ based MOS device. The measurement frequency was 100 kHz at a ramp rate of 50 mV/s, and the switching time for the one complete hysteresis was 40 s.

The difference in $C$-$V_g$ hysteresis from reverse to forward bias direction (=$C_{rf}$ [=$C_r$-$C_f$]) is one way to estimate the border traps ($\Delta N_{BT}$) [8]. The indexes refer to measurements from accumulation to inversion ($C_r = C_{reverse}$) and inversion to accumulation ($C_f = C_{forward}$). This border traps estimation ($\Delta N_{BT}$) is not similar to the one appearing at the classical Si/$SiO_2$ MOS devices where always a single peaked $\Delta N_{BT}$ curve is observed [8]. Fig. 2 depicts a double peaked curve, with peak '1' appearing at weak depletion and close to flatband region, and peak '2' appearing at accumulation; these two peaks can be attributed to contributions from interface and border traps respectively [10]. The enlarged picture of the $\Delta N_{BT}$ is shown as an 'insert' in the graph. At point 'A' of the $C$-$V_g$ curve we can eventually observe a 'bump', which indicates the contribution of the interface traps to the $C$-$V_g$ measurements. Hence both the bump at 'A' and the peak '1' correspond to interface traps, whereas the second peak corresponds to the border traps just like the Si-$SiO_2$ system [8]. Similar results are also observed in other REOs based MOS devices grown on Ge substrates [15, 17].

As mentioned before, there are two types of border traps, the slow and the fast ones. A slower border trap will be counted in the $C$-$Vg$ measurement as a bulk-oxide trap, unlike a faster one which will be counted as an interface trap. This results to the picture of the two peaks at the $\Delta N_{BT}$ curves.

Fig. 3 shows the Capacitance- Voltage(C-$V_g$), Conductance–Voltage ($G_{p/\omega}$-$V_g$) and the estimation of Border Traps ($\Delta N_{BT}$ vs $V_g$) curves, all in one graph, for the shake of clarity and understanding. Estimation of the border traps density ($\Delta N_{BT}$) is required. A measure of the total effective border trap density ($\Delta N_{BT}$) can be obtained by integrating the absolute value of the difference ($C_{rf}$) between the C–Vg curves, using the expression [8, 15]:

$$\Delta N_{BT} \approx \frac{1}{qA} \int |C_r - C_f| dV \quad \text{------------------------------(1)}$$

where, q is the elementary charge, and A is area of the MOSCAPs. The $G_{c/\omega}$-$V_g$ was subjected to series resistance effect correction [18] and all data are normalized to area. The 'first' peak at the depletion region of the $\Delta N_{BT}$ vs. Vg curve and also the 'peak' of the $G_c/\omega$-Vg curve are one-to-one correlated. The corrected ($G_c/\omega$–Vg) curves are strongly peaked at depletion, representing losses due to the exchange of carriers with interface traps [18].

We observe here that the additional peak in the C-$V_g$ difference hysteresis curve ($C_{rf}$) hinters the interface traps contributions. This effect is known as "screening" and it is common in dielectrics other than $SiO_2$ on Si.[9,18]

Fig. 4 shows another experimental fact and convolution the border traps with respect to the progress of time at CVS. If we itemize the evaluation of the $N_{BT}$ under the CVS measurements it is noticeable that with respect to CVS biasing and with progressing stress time, the border traps continue to accumulate and the shape of the $C_{rf}/q$-$V_g$ curves is changing dramatically, suggesting that the total amount of $N_{BT}$ is increasing. This is well illustrated in Fig. 4 where the black solid-square line comes from fresh devices and the red open-circle line is the $\Delta N_{BT}$ after stressing the device at a CVS of -2V for $t_{stress}$=500s . The two peaks 'A' and 'B' in this case, represent the interface traps and border traps contributions to the $C$-$V_g$ hysteresis curve [15] mentioned earlier. It is clear that during the stress conditions new defects are created continuously. Analyzing the above results, the total calculated amount of $N_{BT}$ on fresh and on after stressed devices are equal to $3.1\times10^{12}$eV$^{-1}$cm$^{-2}$ and $3.96\times10^{12}$eV$^{-1}$cm$^{-2}$ respectively. It is important to mention here, that due to the difficulty to distinguish between the border and the interface traps, different groups have reported various opinions and procedures [4,9,18-20], on this issue. These opinions refer either to the contribution of the surface passivation which may alter the physical nature of the defects, or to the fact that the defects density depends on the oxide processing, or both. The REOs are strongly reactive with Ge and during the deposition, the Ge molecules diffuse into it and this intermixing of the Ge and REOs could result to the above mentioned facts [16]. Earlier work on the passivating properties of REOs films showed better electrical quality for the $La_2O_3$ as compared to $CeO_2$ and $Dy_2O_3$ [16, 21], and it is reported that when the interfacial layer of $La_2O_3$ was about 1nm then the $La_2O_3$ or its gate stack ($HfO_2$/$La_2O_3$) didn't demonstrate any additional bump in the $C$-$Vg$ hysteresis measurement compared with those of $CeO_2$, $Dy_2O_3$, and also their gate stacks [22].

**iii) Interface Traps, Border Traps, and Oxides Traps**

$C$-$V_g$ measurements at various CVS bias conditions and frequencies were performed in order to estimate the oxide ($N_{ox}$), interface ($N_{it}$) and effective total border traps ($N_{BT}$) densities in different thicknesses REOs gate stacks (REOs/$HfO_2$), and the results are shown in Figs. 5(a) and (b) for low (-2V) and high (-4V) CVS biases and for 10 consecutive stresses of 500s each (total $t_{stress}$=5000s) respectively. At low CVS the interface traps density is increasing almost exponentially with respect to stress time but the border traps and oxide traps densities remain almost unaltered. The interface traps density is about one order of magnitude higher than the densities of the border and oxide traps.

This increase of the interface traps density at lower CVS which could hinter the positive $\Delta V_{FB}$ shift (see Fig.1a) also contributes to the $\Delta N_{BT}$ curves ambiguity (Figs.2 and 3). At higher CVS conditions the device behavior is different but as usual characteristics (see Fig. 1b) of MOS devices.

The oxide traps are increasing almost one order of magnitude, than the interface and the border traps, which is in agreement with the results shown in Fig.1b. The oxide traps create fixed defects in the bulk, which contribute to the large hysteresis in the $C$-$V_g$ measurements appearing in Fig.1. At higher biases the interface traps density ($N_{it}$) is also increasing but its value remains much lower than that of $N_{ox}$. In both cases, the border traps remain either flat or slightly increasing.

Similar results [22] have been reported in the past for $CeO_2$/Ge MOS devices with anomalous charge trapping, in same polarity stressing, due to the creation of new interface defects at low CVS. This is the distinct feature of the REOs grown on Ge substrates over Si-based MOS devices.

The $N_{it}$ was calculated using a Ge-based simulator (MISFIT [23]) which solves both Poisson and Schrödinger equations simultaneously, taking into account quantum confinement effects, however the border traps were calculated according to Eq.1 and oxide traps by using the methodology of reference [22]. The extraction of the $N_{it}$ from the high frequency $C$-$V$-$G$ measurements of the Ge-based MOS devices is not an easy task, alike that of Si based MOS devices, when the conductance method is used. Batude *et al.* reported that this calculation overestimates $N_{it}$ by almost one order of magnitude in Ge-based MOS capacitors [24], however Bellenger *et al.* suggested contrary [25].

**iv) *J-t* decay transients: Current instabilities**

The use of REOs as a buffer interfacial layer (*il*) demonstrates better passivation and electrical properties compare to other *il* layers between high-*k* (e.g. $HfO_2$, $ZrO_2$) and Ge surface itself [6, 16, 26-27]. However, in terms of reliability, since when gate stacks of high-*κ* dielectrics are used in MOS devices they produce current decay behaviour, (decay transient of *Jg-t*) which is defined as Maxwell-Wagner instabilities (M-W), and many recent reports appear on that in literature [7, 13, 28].

This M-W model, can explain the experimental results (*Jg-t*) with certain limitations (a) until a certain stress time (as $t_{stress} \leq 100s$) (b) when the M-W current ($J_{MW}$) is very low and dominated mainly by the so called Curie von-Schweilder (C-S) relaxation current c) at low CVS regime [7]. On the contrary, at low-medium to medium and certainly at higher CVS, new neutral defects/traps will be created [5, 29-31] which gives rise to stress-induced leakage current (SILC) which is not included in this model. The creation of the neutral defects is defined as

$$J_{SILC} = \alpha . t^{\nu} \quad \text{-----------------------------------------------(2)}$$

where $\alpha$ is the pre-factor which has the dimension of current (A/cm$^2$), and the power $\nu$ is the trap generation rate under bias condition. Therefore if we combine these $J_{M-W}$ and $J_{SILC}$ components they result to a total external circuit current as:

$$J_{(MW,SILC)} = J_{MW} + J_{SILC} \quad \text{--------------------------------- (3)}$$

$$J_{(MW,SILC)} = 2E_{h-k}\sigma_{0,1}\left(3+\ln\frac{t}{t_{0,1}}\right)\frac{t_{0,1}}{t} + \alpha.t^{\nu}, \quad t > t_{0,1} \quad \text{---------(4)}$$

where $E_{h-k}$ is the field across the main high-k dielectrics (here the $HfO_2$, so $E_{HfO2}$), $\sigma_{0,1}$ and $t_{0,1}$ are material constants which have the dimension of conductivity (A/cm$^2$) and relaxation time distribution (s) respectively. The $t_{0,1}$ is expected to be of the order of microsecond to picoseconds [7,13,28].

Using Eq.4, a best fit to the experimental data ($J_g$-$t$) is provided and clearly explains the results [7], see Fig.6. In the literature, the decay transients ($J_g$-$t$) have been described by a field lowering model due to charge trapping at the traps near the gate [33], or by a model using the C-S dielectric relaxation mechanism [34-35,] however the more foreseeable accepted physics explanation is that of the M-W mechanism [7,13,28]. In our case, the proposed model for the current instabilities explains very well the experimental results.

**v) Time dependent dielectrics (gate stacks) degradations**

Finally the devices on both *p*- and *n*- type Ge substrates have been subjected to very long CVS conditions at moderate gate voltages. Fig. 7(a) shows the results for devices grown on *p*-type Ge-substrates and subjected to CVS at $V_g$ = -3.0V ($E_{HfO2}$=3.3MV/cm, $E_{Dy2O3}$=5.9MV/cm). Initially the fit (C-S relaxation $J \sim t^{-n}$) to the $J_g$-$t$ decay (first part) gives the *n* values as *n*~0.56 while the charge trapping is considered for best fitting (Nigam model) [15] of the experimental data, shown in the second part, rising transient of Fig. 7a, where the time constant, $\tau$ was found to be 260s. This device reached hard breakdown after a number of soft breakdown events and a total stress time in the range between 15000s and 20000s. This can be explained such that one of the oxides (probably the thinner, 2nm $Dy_2O_3$) goes to breakdown first, leading to a major redistribution of the corresponding fields. Thus the field across the other dielectric ($HfO_2$) increases abruptly leading to a second relaxation effect. Hence, soft breakdown (SBD) effects appear which, eventually, lead to a hard breakdown (HBD) of the second layer and the device itself. On the contrary, the single Dy$_2$O$_3$ layer (not shown here) needed also a considerably longer time in order to collapse which is probably due to the different breakdown mechanisms of the two oxides.

On the other hand, the application of very long CVS pulses on similar gate stacks (10nm HfO$_2$/1nm Dy$_2$O$_3$) but grown on *n*-type Ge substrates [see Fig.7 b] showed improved reliability characteristics. In particular, at moderate stress fields, (e.g. CVS @ $V_g$= 2.2V where $E_{HfO2}$=1.9MV/cm and $E_{Dy2O3}$=3.3MV/cm at $t_{stress}$=0s) it takes a very long time ($t$=384000s i.e. 4.4 days) in order to observe breakdown characteristics. However these gate stacks show similar behavior to their *p*-type substrates counterparts in the case of the $J_g$-$t$ analysis. In that case, initially the current density decreases due to relaxation effects (C-S relaxation) for 6 seconds, followed by a negative charge trapping (Nigam's Model) in the oxides. When the $J_g$-$t$ increasing transient analyzed by charge trapping model [15], the time constant $\tau$ for charge trapping (second part of the transient) was found to be 47s signifying the presence of neutral traps [15, 32] in the high-$\kappa$ materials. Moreover, the *n* value was calculated to be equal to unity, indicating the unimportance of M-W instabilities here. Therefore, a comparison of the results for the same gate stack configuration on both types of Ge-substrates illustrates the better quality of the *n*-type substrates in terms of electrical reliability. The superior quality of the devices grown on *n*-Ge substrates is in agreement with the well known problem of the *p*-Germanium surface properties as reported by many groups recently [36]. The latter, combined with the results presented in the previous sections, conclude that the use of *n*-Ge substrates is suggested for better quality electrical characteristics for gate stack devices.

## 4. CONCLUSION

The dielectric degradations and electrical reliability characteristics in rare-earth oxides gate stacks on high mobility Ge based devices were studied by means of electrical measurements under CVS. Varying the CVS conditions (low to higher), trapping effects are observed for the same polarity at accumulation condition of the devices which is secernated from other $SiO_2$/Si systems. From border traps analysis we observed the double peaks structure at the $\Delta N_{BT}$ curves which correspond to interface and border traps and they were also verified by complementary electrical measurements ($C$-$G_{p/\omega}$-$V_g$). The contribution of the interface traps at low bias is dominant to the other traps and influences the device degradation. We successfully proposed a Maxwell-Wagner (M-W) mechanism in order to explain the decay current ($J_g$-$t$ transient) at low to higher bias and it was. The mechanism for the breakdown of the gate REOs based stacks is also proposed using the assumption of MW instabilities and progressive breakdown and finally the REOs degrades by hard breakdown (HBD). The different time constants of the charge trapping e.g. 47s and 260s in the degradation of the gate stack based on $n$- and $p$-type Ge substrates respectively indicate the different nature of the defects prevailing in the devices. The previous M-W model has constrains that it is unable to explain the experimental data of longer $Jg$-$t$ transient, also the temperature field dependent as well as the frequency domain behaviour of the relaxation parameters are not included into it. We are currently working on the further development of the modified M-W Model theoretically and experimentally for the gate stacks grown on high mobility substrates.

**FIGURE CAPTIONS**

**Fig. 1(a, b)** (**online color**) High frequency $C$-$V_g$ ($f$ =100 kHz) curves on fresh and stressed devices of Pt/Dy$_2$O$_3$/$p$-Ge. Only the curves fresh and after the application of ten consecutive CVS cycles (500s each) are plotted for clarity. Stress voltage is low in (a) and moderate in (b). Positive $V_{FB}$ shifts in (a) indicate trapping of electron in the bulk of the oxides while negative $V_{FB}$ shifts in (b) indicate creation of positively charged defects.

**Fig. 2** (**online color**) $C$-$V_g$ curve (circled symbol-line) at higher frequency (100kHz) and its hysteresis difference ($C_{rf}$=$C_r$-$C_f$) characteristics (solid line) of a CeO$_2$ based MOS devices. The double-peaked $C_{rf}$ structure corresponds to interface and border traps at deep depletion and accumulation bias region respectively, and the enlarged double-peaked curve is shown as insert in the figure for clarity.

**Fig. 3** (**online color**) shows the Capacitance- Voltage (C-$Vg$), Conductance–Voltage ($G_{p/}$ -$V_g$) and the estimation of Border Traps ( $N_{BT}$ vs $V_g$) curves, all in one graph. The bump in the $C$-$V_g$ curve (because of interface defects) and the $G_p$/ peak (representing the losses due to the exchange of carriers with interface traps) show one-to-one relation with the second peak of the $C_{rf}$ curve. Thus graph also confirms the additional peak due to $N_{it}$ contribution to the $C_{rf}$ curve from the Border traps analysis.

**Fig. 4** (**online color**) illustrates the evolution of 'border traps' in Dy$_2$O$_3$/HfO$_2$ gate stacks under constant voltage stress (CVS) conditions. Total number of 'Border traps'

increases with the progress of time (stress time, 500s) during CVS in gate stacks which are equal to $3.1\times10^{12}eV^{-1}cm^{-2}$ and $3.96\times10^{12}eV^{-1}cm^{-2}$ respectively.

**Fig. 5 (a,b) (online color)** The evolution of Oxide traps ($N_{ox}$), Border traps ($N_{BT}$), and interface traps ($N_{it}$) of Dy$_2$O$_3$/HfO$_2$ gate stacks at different CVS of low(-2V) and moderate (-4V) under 10 successive stresses of 500s each respectively. At low CVS the $N_{it}$ density is increasing almost exponentially and one order of magnitude higher with respect to $N_{ox}$ and $N_{BT}$ in stress time while at higher bias $N_{ox}$ shows the similar behavior i.e. $N_{ox}$ is increasing. These results also support with the phenomena are observed in Fig. 1(a) and (b) respectively.

**Fig. 6 (online color)** shows gate current as a function of stress time of gate stack (HfO$_2$/Dy$_2$O$_3$) semi-log plot. The dotted (blue) and solid (red) lines are fits according to Eq (3) and (4) to the experimental data. The proposed model (4) for current instability ($J_g$-$t$ transient) the M-W effects together with SILC eplains completely the experimental data while the previous model (3) unable to fit the data completely (only initial first 40s).

**Fig. 7 (online color)** Current density $J_g$ as a function of stress time $t$ for samples of HO$_2$/Dy$_2$O$_3$ gate stacks grown on *n*- and *p*- type Ge substrates at moderate CVS conditions applied for very long times. The device shows initially relaxation behavior, then (from $t$>12s for *n*-type, $t$>90s for *p*-type) charge trapping to the preexisting traps in the oxide takes place which eventually leads to breakdown.

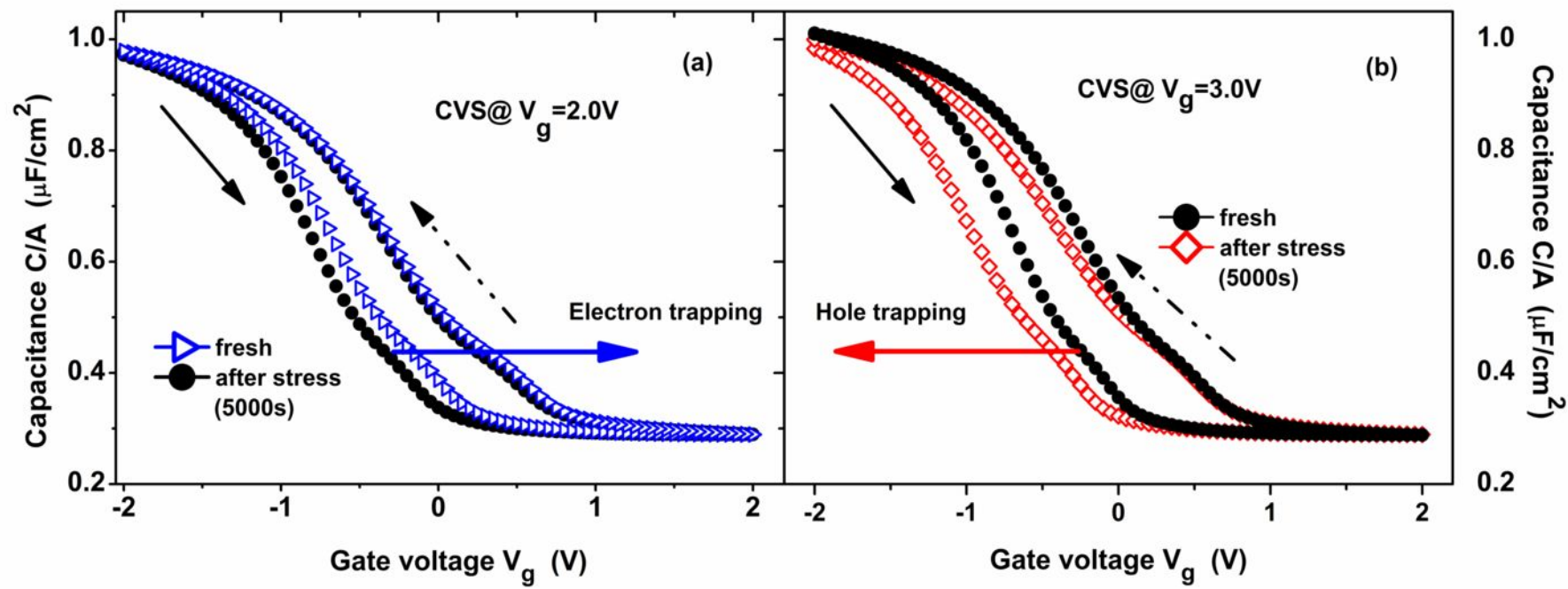

Fig.1

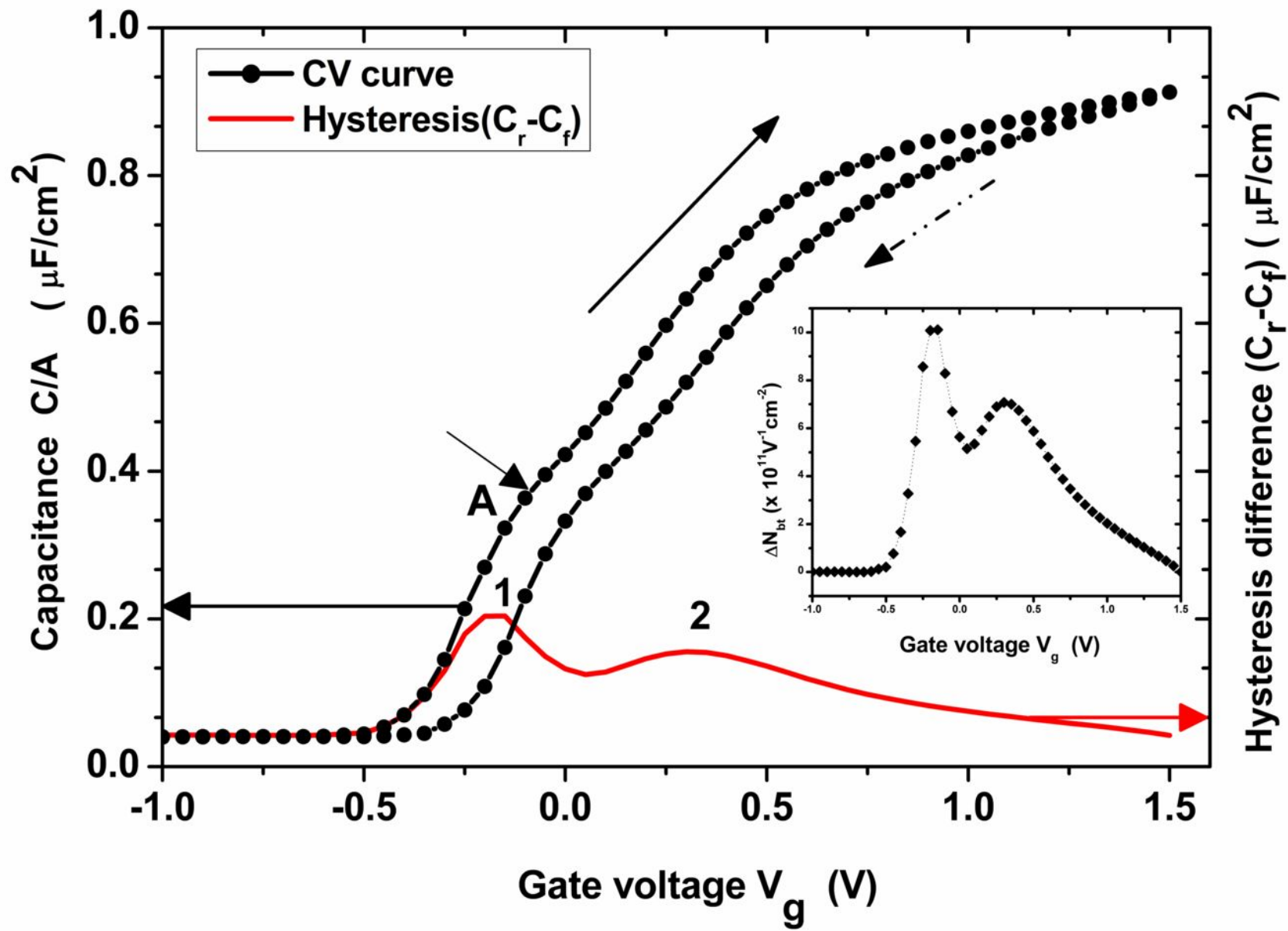

Fig.2

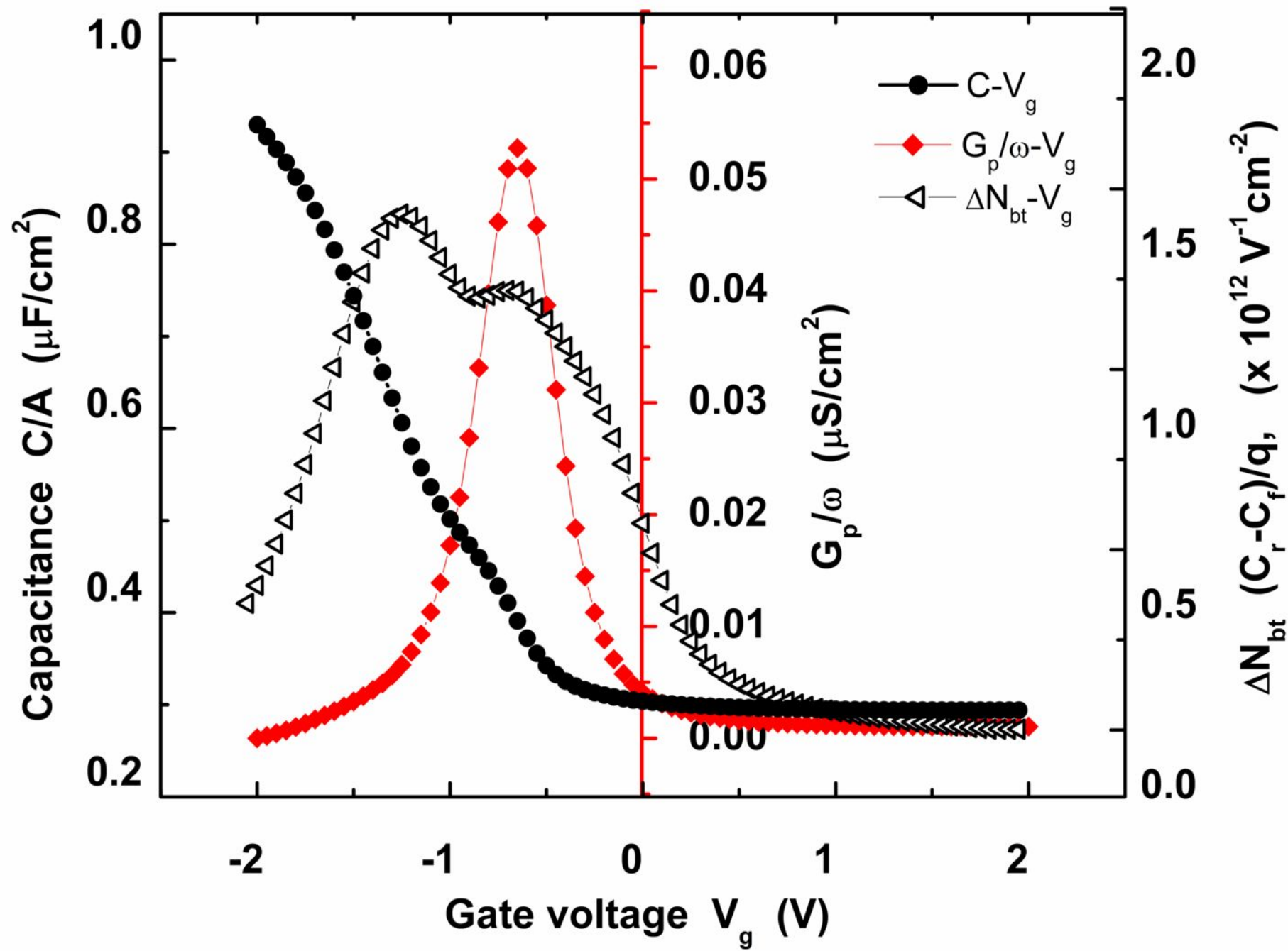

Fig.3

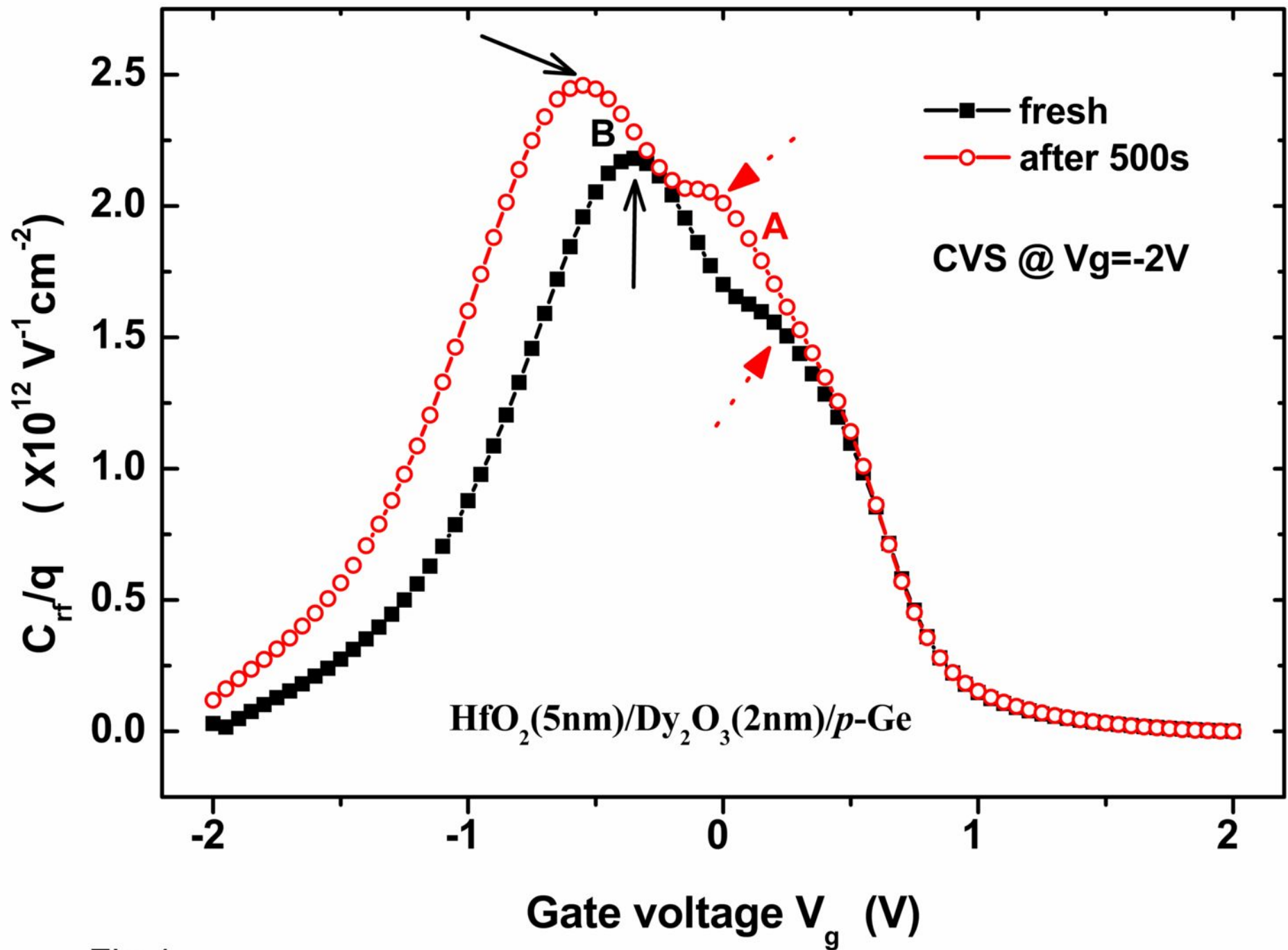

Fig.4

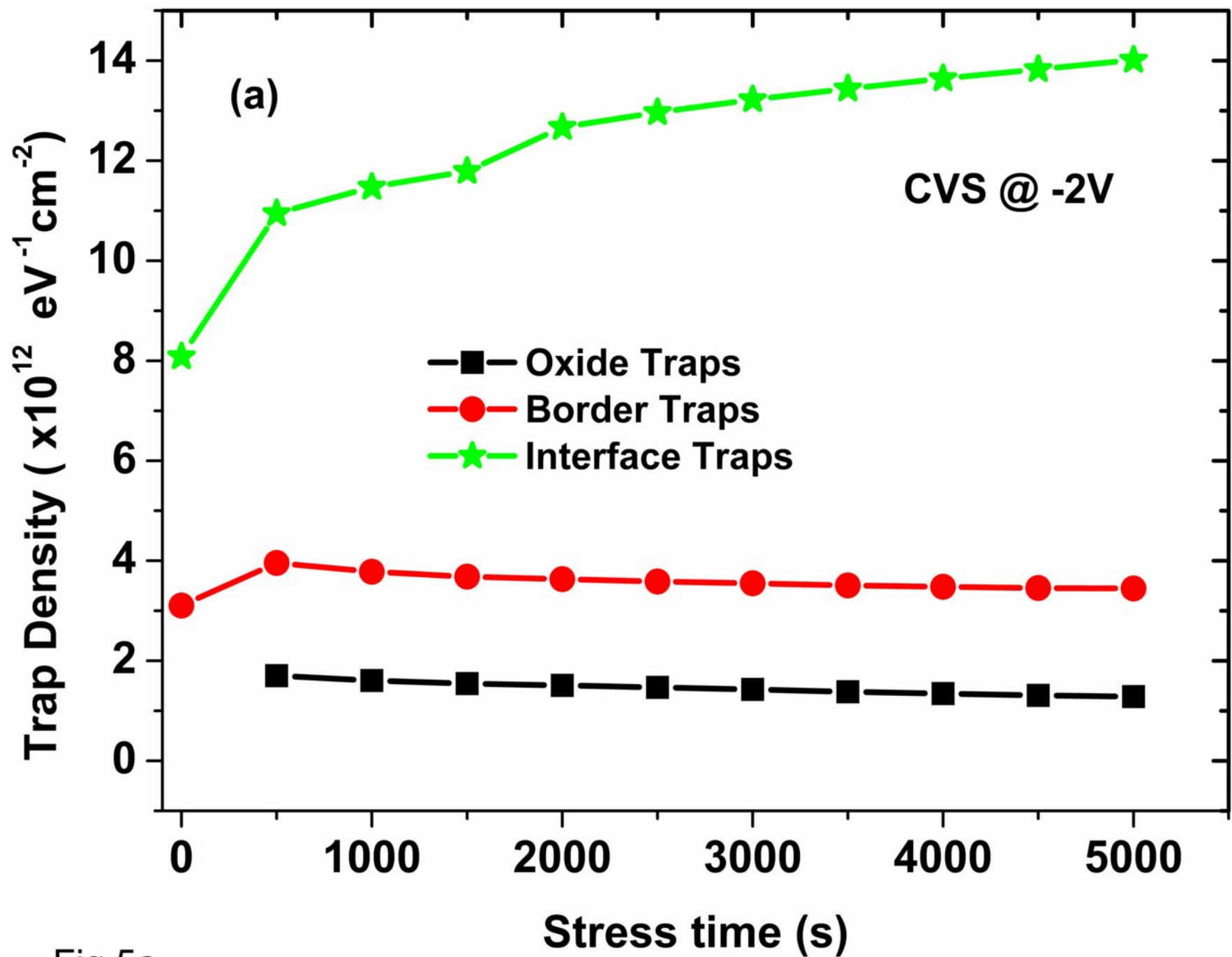

Fig.5a

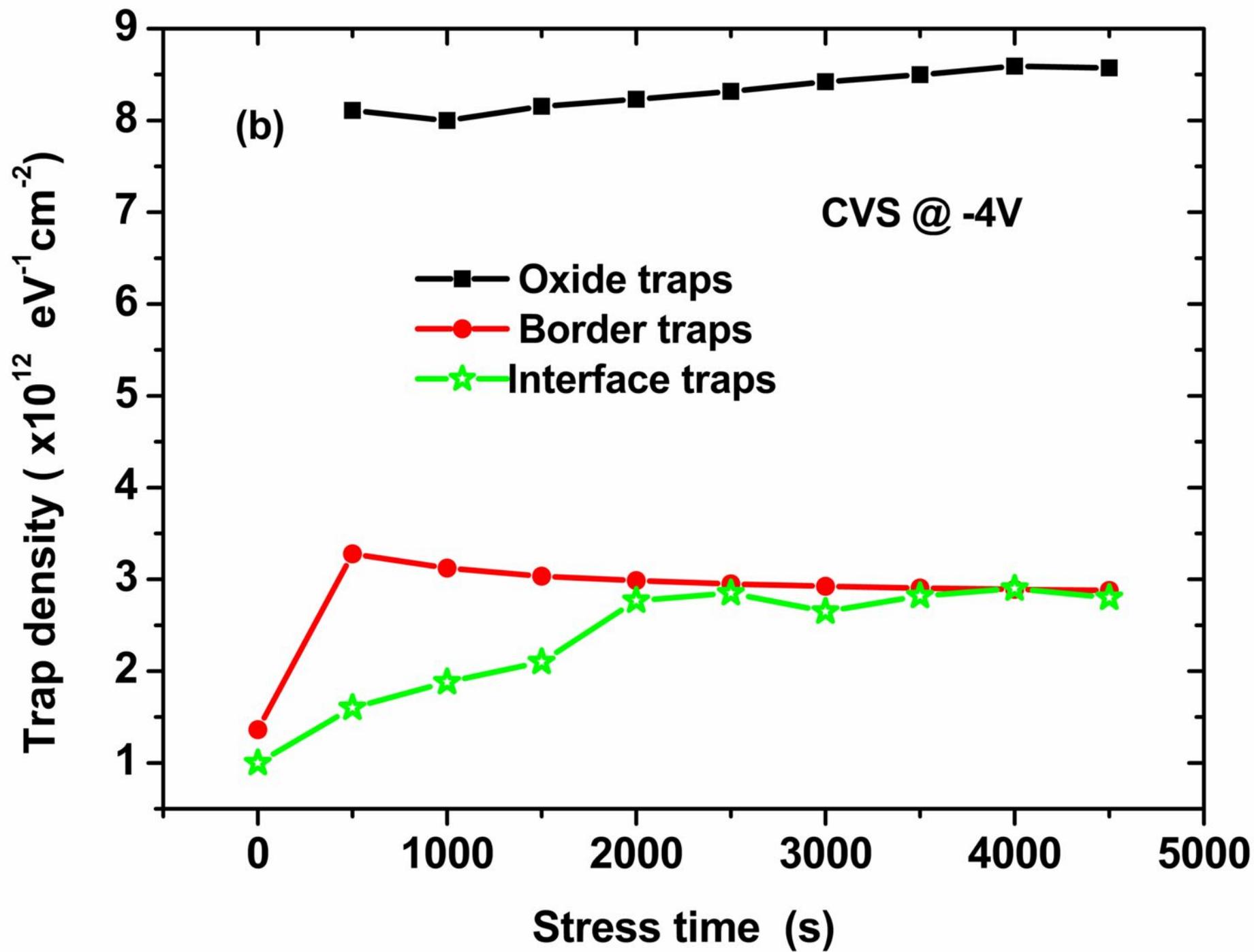

Fig5b

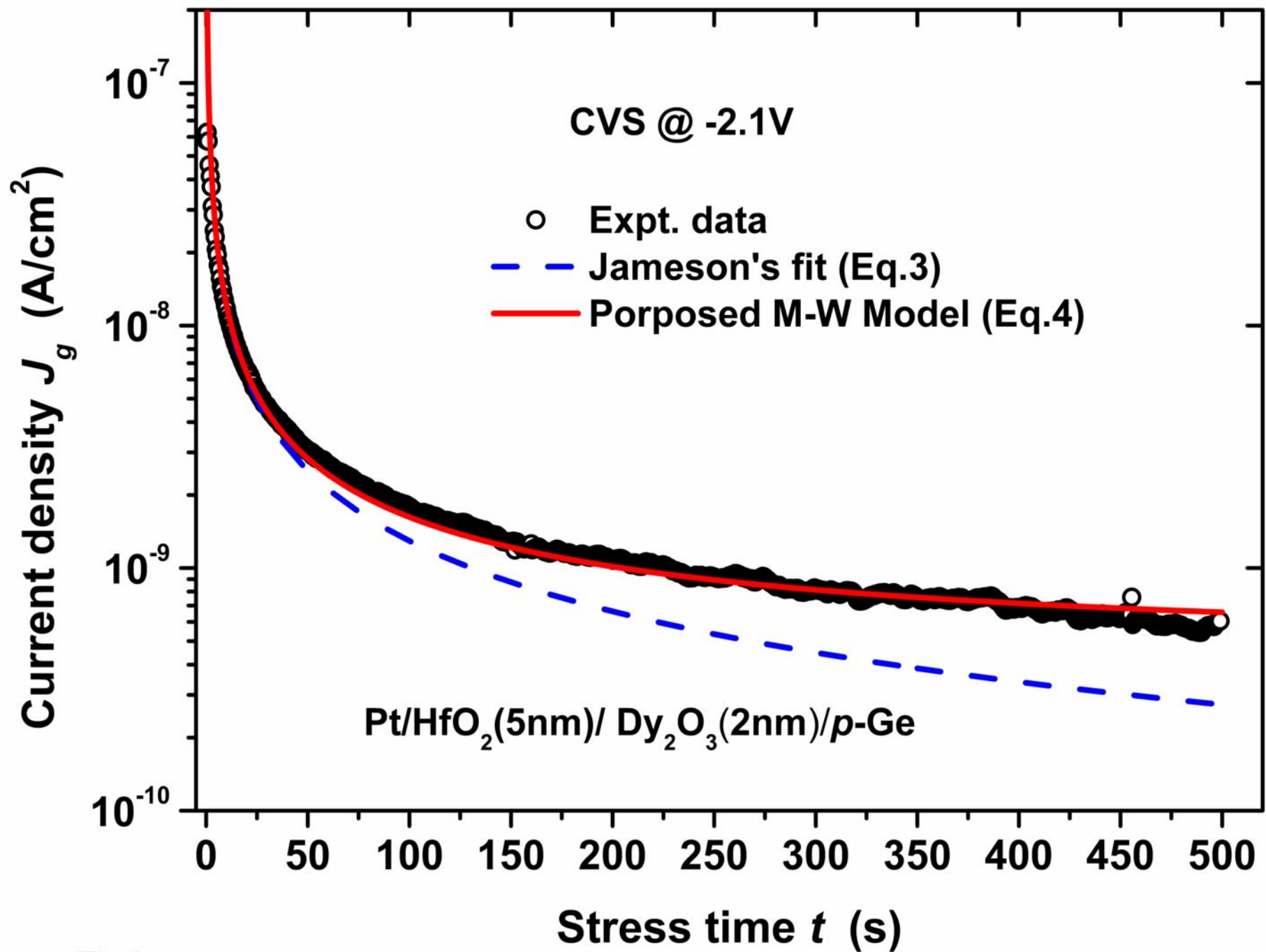

Fig6

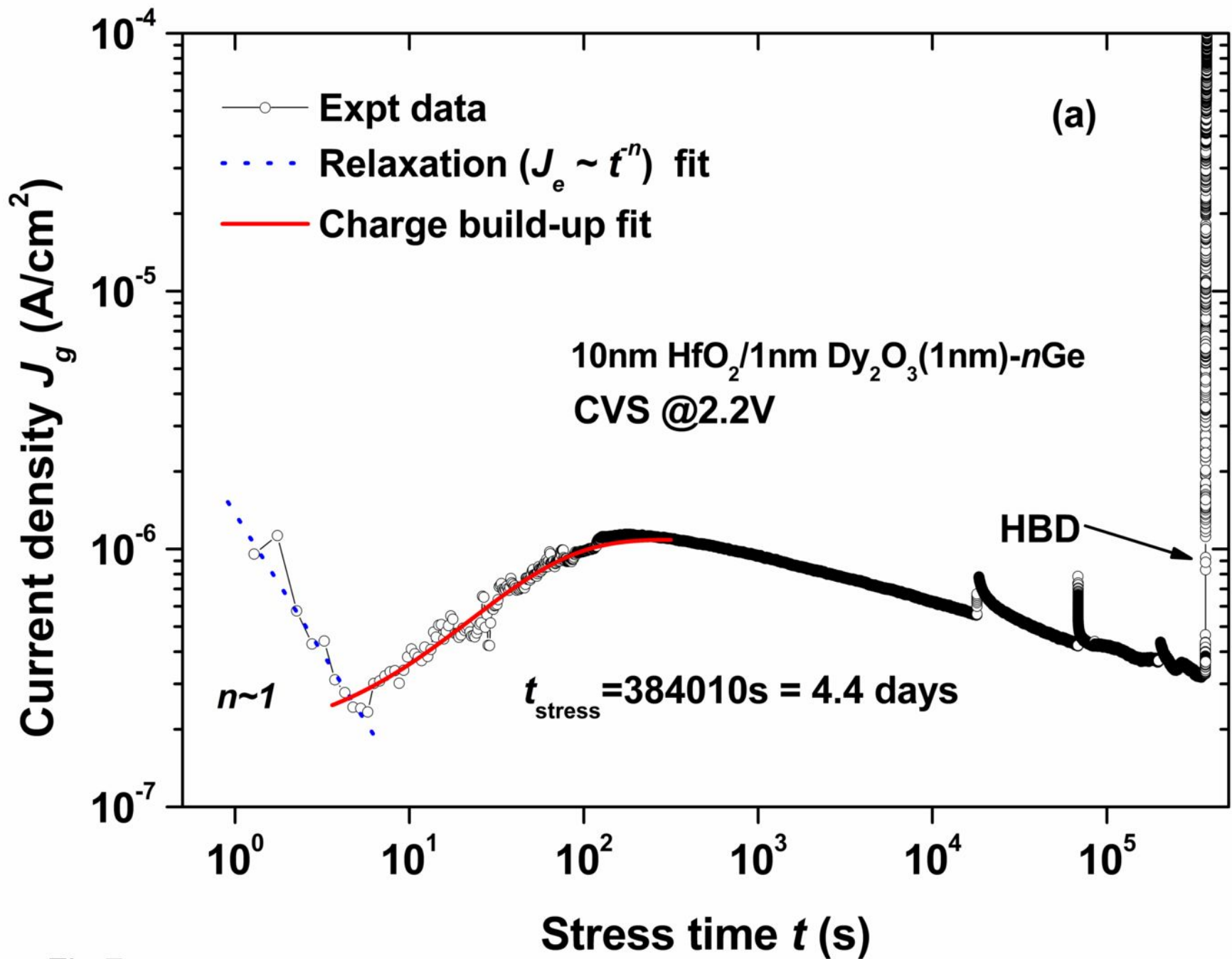

Fig.7a

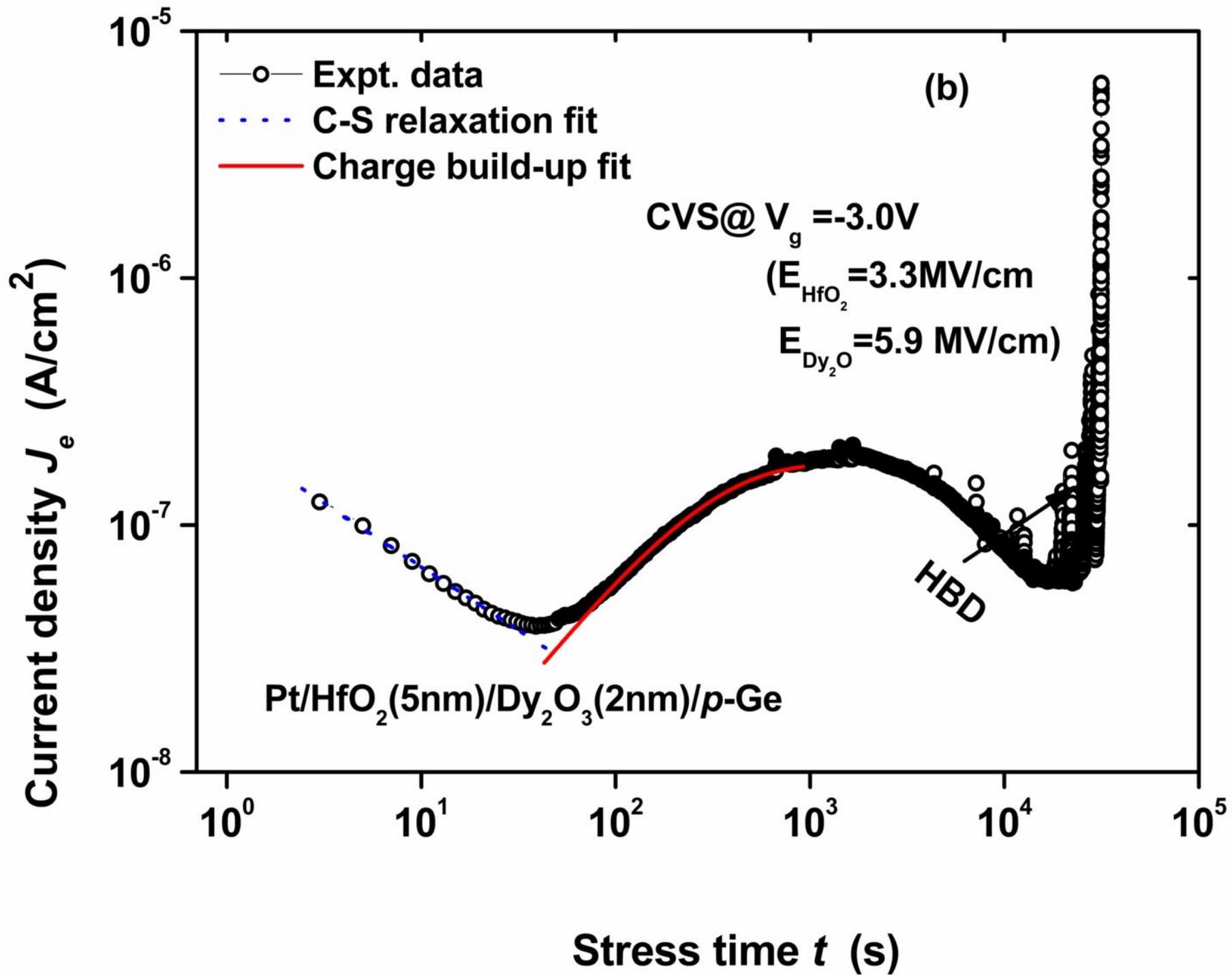
Fig7b